\documentclass[letterpaper, 10 pt, conference]{ieeeconf}  

\IEEEoverridecommandlockouts                              

\usepackage{amsmath}
\usepackage{amsfonts}
\usepackage{booktabs}
\usepackage{graphicx}

\title{\LARGE \bf
Linkage Attacks Expose Identity Risks in Public ECG Data Sharing
}

\author{
    Ziyu Wang$^{*1}$, Elahe Khatibi$^{*1}$, Farshad Firouzi$^{2}$,\\
    Sanaz Rahimi Mousavi$^{3}$, Krishnendu Chakrabarty$^{2}$, Amir M. Rahmani$^{1}$
    \thanks{*Ziyu Wang and Elahe Khatibi contributed equally to this work.}
    \thanks{$^{1}$ University of California, Irvine, USA. 
        {\tt\small \{ziyuw31, ekhatibi, a.rahmani\}@uci.edu}}
    \thanks{$^{2}$ Arizona State University, USA. 
        {\tt\small \{Farshad.Firouzi, Krishnendu.Chakrabarty\}@asu.edu}}
    \thanks{$^{3}$ California State University, Dominguez Hills, USA. 
        {\tt\small srahimimoosavi@csudh.edu}}
}

\begin{document}

\maketitle


\begin{abstract}
The increasing availability of publicly shared electrocardiogram (ECG) data raises critical privacy concerns, as its biometric properties make individuals vulnerable to linkage attacks. Unlike prior studies that assume idealized adversarial capabilities, we evaluate ECG privacy risks under realistic conditions where attackers operate with partial knowledge. Using data from 109 participants across diverse real-world datasets, our approach achieves 85\% accuracy in re-identifying individuals in public datasets while maintaining a 14.2\% overall misclassification rate at an optimal confidence threshold, with 15.6\% of unknown individuals misclassified as known and 12.8\% of known individuals misclassified as unknown. These results highlight the inadequacy of simple anonymization techniques in preventing re-identification, demonstrating that even limited adversarial knowledge enables effective identity linkage. Our findings underscore the urgent need for privacy-preserving strategies, such as differential privacy, access control, and encrypted computation, to mitigate re-identification risks while ensuring the utility of shared biosignal data in healthcare applications.
\end{abstract}


\section{Introduction}

Electrocardiograms (ECG) capture the heart’s electrical activity, serving as a key diagnostic tool for conditions like heart failure and arrhythmias~\cite{schlapfer2017computer, aqajari2024enhancing}. Beyond clinical use, ECG signals exhibit unique morphological and temporal patterns influenced by heart anatomy, conduction pathways, and autonomic regulation, making them suitable for biometric identification~\cite{ghazarian2021increased}. Unlike fingerprints, ECG signals are dynamic and vary naturally due to physiological changes, enabling intrinsic condition detection, which reduces susceptibility to spoofing and enhances security in authentication systems~\cite{odinaka2012ecg}. As telehealth platforms and wearable devices integrate ECG data for remote monitoring~\cite{alikhani2024seal, alikhani2024ea}, the increasing availability of publicly shared datasets raises significant privacy concerns~\cite{wang2024ecg}.

Due to their biometric nature, ECG signals are vulnerable to linkage attacks, where adversaries exploit overlaps between public and private datasets to re-identify individuals. Unlike demographic data, ECG retains distinctive patterns over time, making de-identification challenging. An attacker with partial knowledge can aggregate ECG samples from wearables, telehealth platforms, or leaked records and cross-reference them with public datasets using machine learning~\cite{biel2001ecg}. Leveraging membership inference~\cite{shokri2017membership} and record linkage techniques, attackers can link anonymized ECGs to identities, compromising patient privacy. This risk grows as ECG datasets proliferate across research, clinical, and commercial domains without adequate safeguards.

\begin{figure}[t]
    \centering
    \includegraphics[width=0.9\linewidth]{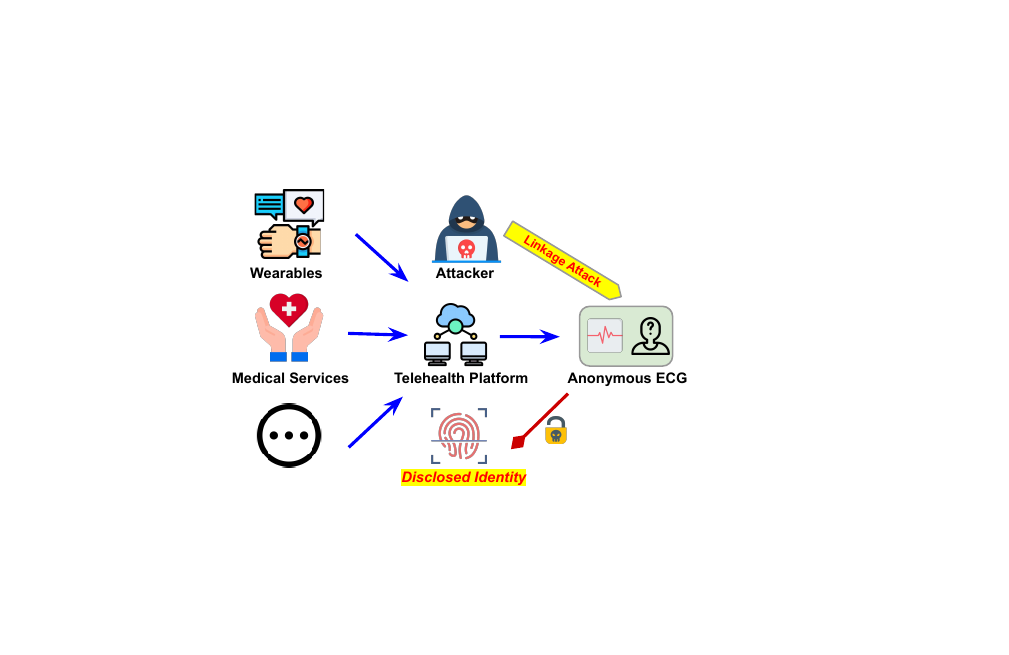}
    \caption{Illustration of a linkage attack on public anonymous ECG data. A telehealth platform integrates data from multiple sources, including wearable devices and medical institutions. An attacker with partial knowledge cross-references public and private datasets to re-identify individuals, compromising privacy and trust in medical data sharing.}
    \label{fig:linkage_attack}
\end{figure}

Figure~\ref{fig:linkage_attack} demonstrates this problem, showing how telehealth platforms consolidate ECG data from diverse sources, creating a rich repository that attackers can exploit~\cite{wang2020guardhealth, yao2020privacy}. By correlating leaked or externally obtained ECG signals with public datasets, adversaries can disclose identities, leading to privacy breaches. Beyond compromising confidentiality, linkage attacks undermine trust in medical data sharing, discourage research participation, and threaten the security of digital health infrastructures~\cite{wang2024differential}. As healthcare increasingly relies on open data initiatives, mitigating these risks is essential to preserving patient privacy and trust~\cite{wang2024ecg, wang2025transecg}.

Despite these risks, existing ECG privacy studies often make overly simplified assumptions about adversarial capabilities~\cite{labati2019deep}, assuming full data access or controlled re-identification settings. In reality, attackers operate with partial and noisy knowledge, handling heterogeneous datasets from varying devices, medical conditions, and recording environments. Moreover, prior studies struggle to distinguish known from unknown individuals~\cite{ghazarian2021increased, pelc2019ecg}, inflating success rates in controlled settings while overlooking real-world challenges like within-person variation due to aging, disease progression, or sensor differences~\cite{wang2025healthq}. These gaps hinder a true understanding of ECG privacy risks and the development of effective protective measures.

To address ECG privacy risks, we propose the first framework to assess re-identification under realistic linkage attacks. Unlike prior studies assuming full adversarial knowledge, we model a practical threat where attackers with partial knowledge cross-reference auxiliary ECG data with public repositories to link identities, simulating real-world risks from wearables, telehealth, and leaked medical records. We implement a Vision Transformer (ViT)-based model~\cite{dosovitskiy2020image} to assess how well attackers can match ECG signals while correctly distinguishing unknown individuals. Our study provides an accurate evaluation of linkage attack risks in biosignal data sharing, demonstrating the urgent need for stronger privacy-preserving measures in healthcare.

\textbf{Our contributions are as follows:}  
\begin{itemize}  
    \item We define and formalize ECG linkage attacks, illustrating how adversaries exploit overlaps between public and private datasets for re-identification.  
    \item We introduce the first realistic attack model that quantifies re-identification risks under partial adversarial knowledge, aligning with real-world telehealth and data-sharing constraints.  
    \item We provide empirical evidence of ECG privacy vulnerabilities across diverse datasets, establishing benchmarks to guide privacy-preserving research.  
\end{itemize}

\section{Methodology}

\begin{figure*}[htbp]
    \centering
    \includegraphics[width=0.8\linewidth]{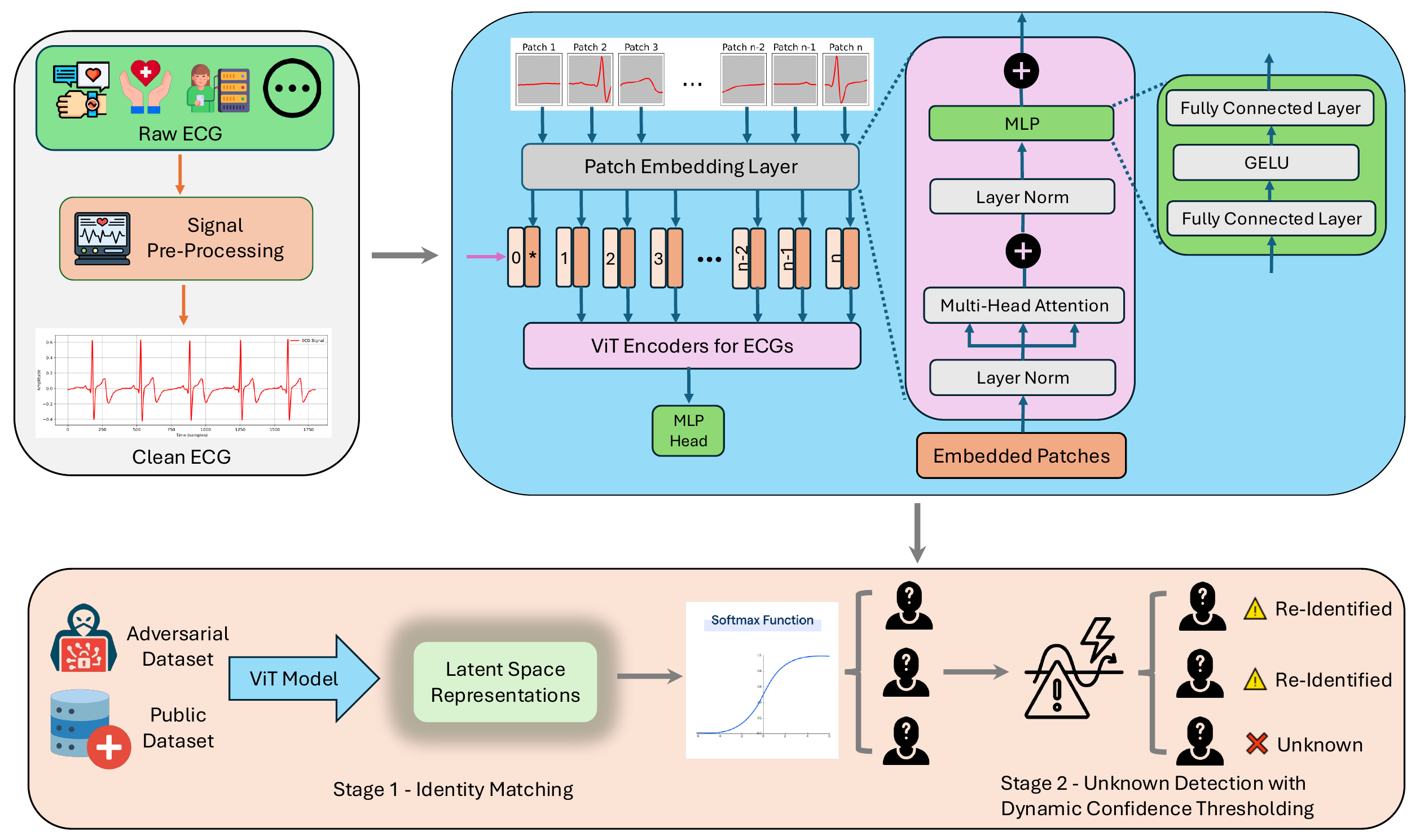}
    \caption{Overview of the proposed linkage attack method.}
    \label{fig:model}
    \vspace{-1.3em}
\end{figure*}

The biometric nature of ECG signals makes publicly shared datasets vulnerable to linkage attacks, where adversaries exploit overlaps between public and private data to re-identify individuals. Figure~\ref{fig:model} illustrates our ViT-based framework. First, raw ECG signals undergo pre-processing, segmentation, and embedding. A ViT encoder extracts identity-specific features, followed by an Multi-Layer Perceptron (MLP) classifier. Stage 1 (Identity Matching) leverages an adversarial dataset—collected from public or leaked sources—to train a ViT model that maps ECG signals to latent space representations, assigning identity probabilities via softmax. Stage 2 (Unknown Detection) applies a dynamic confidence threshold to reject uncertain matches, reducing false identifications. This framework quantifies re-identification risks, showing that adversaries can link ECG data even with partial knowledge, underscoring the need for stronger privacy safeguards.




\subsection{Problem Definition}

Let $\mathcal{D}_{\text{pub}} = \{(x_i, y_i)\}_{i=1}^{N}$ be a public ECG dataset, where $x_i \in \mathbb{R}^T$ is an ECG signal of length $T$, and $y_i \in \mathcal{Y}$ is the corresponding identity label. Similarly, let $\mathcal{D}_{\text{priv}} = \{x_j\}_{j=1}^{M}$ be a private dataset containing ECG signals from both \textit{known participants}, whose identities exist in $\mathcal{D}_{\text{pub}}$, and \textit{unknown participants}, who are absent from $\mathcal{D}_{\text{pub}}$. Given $x_j \in \mathcal{D}_{\text{priv}}$, an attacker seeks to assign a label $\hat{y}_j \in \mathcal{Y} \cup \{\text{unknown}\}$, identifying whether the signal belongs to a known or unseen individual.

We investigate linkage attack risks, where an adversary attempts to re-identify individuals by cross-referencing ECG samples in $\mathcal{D}_{\text{priv}}$ with identities in $\mathcal{D}_{\text{pub}}$. The attacker uses an adversarial dataset—ECG signals collected from public sources, leaked medical data, or unauthorized means, containing identifiable records. Unlike standard classification, the attacker operates under partial knowledge, meaning they do not know whether a given identity exists in $\mathcal{D}_{\text{pub}}$. This introduces practical challenges, such as dataset mismatches, ECG variability across devices and conditions, and the need to differentiate between known and unknown individuals. Our study formalizes this threat model and quantifies the re-identification risks posed by such attacks in real-world data-sharing scenarios.

\subsection{Framework for Linkage Attack Evaluation}

The proposed framework evaluates re-identification risks by simulating an attacker attempting to link ECG samples from $\mathcal{D}_{\text{priv}}$ to individuals in a black-box public dataset $\mathcal{D}_{\text{pub}}$. Unlike traditional attacks that assume full adversarial knowledge~\cite{wang2024ecg, ghazarian2021increased}, our framework simulates real-world constraints, where the attacker has limited visibility and no prior knowledge of the overlap between $\mathcal{D}_{\text{priv}}$ and $\mathcal{D}_{\text{pub}}$. The attacker employs a machine learning-based attack using a ViT, training on an adversarial dataset (ECG data collected from public records, leaks, or unauthorized sources) without explicit knowledge of which identities exist in the target dataset.

\subsection{ViT-Based Representation Learning}

To effectively encode ECG signals, the ViT-based model~\cite{dosovitskiy2020image} applies self-attention mechanisms to capture both local and global dependencies in the signal. Unlike convolutional architectures~\cite{labati2019deep}, which primarily rely on spatial locality, the ViT capture long-range dependencies, making them well-suited for biosignals.

\paragraph{Patch Embedding and Positional Encoding}  
Given an ECG sequence \( x_i \in \mathbb{R}^{T} \) from $\mathcal{D}_{\text{priv}}$, we divide it into fixed-length patches of size \( P \), resulting in \( N = T / P \) patches. Each patch is linearly projected into a \( d \)-dimensional space:  
\[
z_0 = W_p x_i + b_p
\]
where \( W_p \in \mathbb{R}^{d \times P} \) is a learnable projection matrix. A learnable classification token \( z_{\text{cls}} \) is prepended to the sequence, and positional encodings \( E \) are added to retain temporal structure:  
\[
Z = [z_{\text{cls}}, z_0, z_1, ..., z_N] + E
\]
This embedding is passed through multiple transformer layers for feature extraction.

\paragraph{Multi-Head Self-Attention Mechanism}  
Each transformer layer consists of a multi-head self-attention module that computes relationships between input patches across multiple attention heads. The attention scores for each head are computed as:  
\[
A_h = \text{softmax} \left( \frac{Q_h K_h^T}{\sqrt{d_k}} \right)
\]
where \( Q_h, K_h, V_h \) are the query, key, and value matrices for head \( h \), and \( d_k \) is the scaling factor. The outputs from all heads are concatenated and linearly transformed:  
\[
Z' = \text{Concat}(A_1 V_1, A_2 V_2, ..., A_H V_H) W_O
\]
where \( H \) is the number of attention heads, and \( W_O \) is a learnable projection matrix. This is followed by layer normalization and a feedforward network. The final feature representation is extracted from \( z_{\text{cls}} \) and passed to a classifier, encoding the ECG signal into a latent feature space where identity-specific characteristics are captured.

\subsection{Two-Stage Classification for Known and Unknown Participants}

With latent space representations from the ViT, the attacker seeks to match an ECG sample to an identity in $\mathcal{D}_{\text{pub}}$. However, without prior knowledge of the target dataset's composition, direct classification is uncertain, risking false matches for unseen individuals. Notably, this attack remains practical—the attacker trains a model solely on their adversarial dataset, without external knowledge or privileged access. Rather than relying on handcrafted rules, the model autonomously learns patterns, demonstrating that re-identification is feasible even with limited adversarial knowledge.
 
To address this challenge, the attack follows a two-stage classification process:  

\paragraph{Stage 1: Identity Matching}  
The ViT model processes the input ECG sample and outputs a probability distribution over the set of known identities:  
\[
P(y | x_j; \theta) = \text{softmax}(Wz + b),
\]
where \( W \) and \( b \) are trainable parameters. The identity with the highest probability is selected as the predicted label:  
\[
\hat{y}_j = \arg\max P(y | x_j; \theta).
\]
This step assumes that if the sample belongs to an individual already present in $\mathcal{D}_{\text{pub}}$, the model will confidently assign the correct identity. However, if the sample comes from an unknown participant, the model may still attempt to match it to the closest known identity, leading to misclassification.  

\paragraph{Stage 2: Unknown Detection with Dynamic Confidence Threshold}  
Since the attacker lacks full knowledge of $\mathcal{D}_{\text{pub}}$, it is critical to determine when a test sample does not match any known identity. To mitigate false positive matches, we introduce a dynamic confidence threshold \( \Phi \), which adapts based on the confidence scores of known samples:  
\[
\Phi = \text{percentile}(\{\tau(x) \mid x \in \mathcal{D}_{\text{pub}}\}, p),
\]
where \( \tau(x) \) is the confidence score of the highest prediction, and \( p \) is a tunable percentile parameter. This means that the threshold is determined by the confidence levels observed in the public dataset, making it adaptable to different distributions of ECG signals.  

If the model's confidence in its highest prediction for $x_j$ is below \( \Phi \), the sample is classified as \textit{unknown}, rather than being forcefully assigned to a known identity. This approach allows the attack to selectively reject uncertain predictions, reducing the risk of misidentifying individuals who are not in $\mathcal{D}_{\text{pub}}$. By dynamically adjusting \( \Phi \), the model balances sensitivity and specificity, improving the robustness of the attack. Despite these constraints, this attack remains effective because the attacker leverages only their own independently trained model, making no assumptions about external datasets, manually crafted decision rules, or privileged knowledge about $\mathcal{D}_{\text{pub}}$.

\subsection{Design Considerations and Practical Constraints}

Our framework simulates real-world adversarial conditions, ensuring the attack reflects practical constraints rather than an idealized setting. Since ECG signals encode both biometric and physiological variations, the ViT model effectively extracts distinguishing features while handling variability. A dynamic confidence threshold mitigates false matches by preventing identity assignments under high uncertainty. This study provides a realistic assessment of privacy risks in publicly shared ECG data, highlighting the limitations of traditional anonymization techniques like noise addition, which often degrade diagnostic value. Alternative privacy-preserving strategies, such as differential privacy~\cite{wang2024differential} and federated learning~\cite{coelho2023multimodal}, are needed to mitigate re-identification risks while preserving ECG utility.

\section{Experiments and Results}

We conduct a comprehensive evaluation of our framework for ECG linkage attack analysis, focusing on re-identification risks, robustness under adversarial settings, and benchmarking against state-of-the-art methods. Additionally, we analyze the impact of different data splits and perform an ablation study to assess key framework components.

\subsection{Datasets}

To ensure a robust evaluation, we utilize publicly available ECG datasets covering diverse demographics, recording conditions, and health states, summarized in Table~\ref{tab:ecg_datasets}. Each dataset undergoes preprocessing, including resampling to a consistent sampling rate, segmentation into fixed-length windows (e.g., 10s intervals), and z-score normalization.

\begin{table*}[!ht]
\vspace{-8pt}
\scriptsize
\centering
\caption{Summary of ECG Datasets Used in Experiments}
\vspace{-5pt}
\label{tab:ecg_datasets}
\begin{tabular}{|l|c|c|c|c|c|}
\hline
\textbf{Dataset} & \textbf{Subjects} & \textbf{Age Range} & \textbf{Gender (M/F)} & \textbf{Sampling Rate (Hz)} & \textbf{Health Condition} \\ \hline
MIT-BIH Arrhythmia~\cite{moody2001impact} & 47 & 23-89 & 25/22 & 360 & Arrhythmias \\ \hline
MIT-BIH Malignant Ventricular~\cite{greenwald1986development} & 22 & 22-78 & -- & 250 & Ventricular Tachycardia \\ \hline
BIDMC CHF~\cite{baim1986survival} & 15 & 22-71 & 11/4 & 250 & Congestive Heart Failure \\ \hline
Brno University of Tech ECG Quality~\cite{nemcova2020brno} & 15 & 21-83 & 9/6 & 1000 & General Population \\ \hline
MIT-BIH Long-Term~\cite{goldberger2000physiobank} & 7 & 46-88 & 6/1 & 128 & Long-Term General Monitoring \\ \hline
\textbf{Combined} & \textbf{106} & \textbf{21-89} & \textbf{Varies} & \textbf{Varies} & \textbf{Multiple} \\ \hline

\end{tabular}
\end{table*}

\subsection{Experimental Setup}

\paragraph{Adversarial Scenarios}  
To evaluate the robustness of our framework, we simulate realistic attack conditions under different levels of adversarial knowledge. The attacker's goal is to re-identify individuals in the public dataset $\mathcal{D}_{\text{pub}}$ using an adversarial dataset $\mathcal{D}_{\text{priv}}$. However, the attacker does not know in advance whether an individual in $\mathcal{D}_{\text{priv}}$ exists in $\mathcal{D}_{\text{pub}}$. We consider the following scenarios:

\begin{itemize}

    
    \item \textbf{Partial Knowledge (Realistic Scenario):} The attacker trains on $\mathcal{D}_{\text{priv}}$, which includes ECG samples from individuals who may or may not exist in $\mathcal{D}_{\text{pub}}$. Without prior knowledge of dataset overlap, the attacker relies on independently collected data to attempt re-identification, reflecting real-world constraints.
    
    \item \textbf{Full Knowledge (Upper Bound):} The attacker has full access to $\mathcal{D}_{\text{pub}}$ during training, ensuring all target individuals are present. This reduces the task to standard classification rather than true re-identification, representing an unrealistic but best-case scenario for attack performance.

    \item \textbf{Noisy Data (Robustness Test):} To evaluate the framework’s resilience, we introduce synthetic noise into $\mathcal{D}_{\text{pub}}$, simulating real-world distortions such as sensor errors or variations in ECG recordings. This tests the model’s ability to generalize under imperfect data conditions.
\end{itemize}

\paragraph{Data Splits}  
To systematically assess re-identification risks, we design training-validation-testing splits that reflect practical adversarial constraints. The attacker's dataset is constructed by sampling from $\mathcal{D}_{\text{pub}}$, but the attacker does not know in advance which individuals are present in the test set. Specifically, we divide $\mathcal{D}_{\text{pub}}$ into three subsets:

\begin{itemize}
    \item \textbf{Known Individuals:} Participants whose ECG data appears in both $\mathcal{D}_{\text{pub}}$ and the attacker’s training set. The attacker learns to classify these individuals but does not have access to their full ECG sequences.
    \item \textbf{Unknown Individuals:} Participants in $\mathcal{D}_{\text{pub}}$ whose data is completely absent from the attacker’s training set. The attacker must recognize that these individuals are not part of the known set and classify them as \textit{unknown} rather than incorrectly linking them to an existing identity.
    \item \textbf{Test Set:} A mix of known and unknown individuals used to evaluate the model’s ability to correctly match identities while minimizing false positives and false negatives.
\end{itemize}

We employ multiple data splits (e.g., 50/25/25, 60/20/20, and 70/15/15 for training, validation, and testing) to assess how different levels of data exposure influence attack success. These splits simulate a practical attack scenario where the adversary is trained only on a subset of known individuals and must distinguish between known and unknown participants in the test phase. Since the attacker does not have full visibility into the dataset, they must generalize from limited observations, making this a realistic evaluation of re-identification risks in public ECG data sharing.

\paragraph{Evaluation Metrics} 
To comprehensively assess model performance, we employ the following evaluation metrics:

\begin{itemize}
    \item \textbf{Sample-Level Metrics:} Accuracy, precision, recall, and F1-score measure how well individual ECG samples are correctly identified. Higher precision indicates fewer false positives, while higher recall reflects fewer missed identifications. These metrics capture the reliability of ECG-based identity matching.
    
    \item \textbf{Participant-Level Metrics:} Re-identification rate measures how often the model correctly links individuals from the adversarial dataset to the target dataset. Protection rate quantifies how well the system prevents unknown individuals from being misclassified as known identities. These metrics assess identity leakage risks in practical settings.
    
    \item \textbf{Unknown Detection Metrics:} True Negative Rate (TNR) measures the ability to correctly reject unknown individuals, while False Positive Rate (FPR) captures how often they are mistakenly linked to known identities. Confidence thresholds regulate the decision boundary, helping the model avoid overconfident misclassifications in real-world deployments.

    \item \textbf{Threshold-Based Metrics:} Since confidence thresholds critically impact classification performance, we evaluate Equal Error Rate (EER) \cite{nolin2023privecg}, defined as the threshold where the False Acceptance Rate (FAR) equals the False Rejection Rate (FRR):
        \begin{equation*}
        \text{EER} = \arg\min_{\theta} | \text{FAR}(\theta) - \text{FRR}(\theta) |
        \label{eq:eer}
        \end{equation*}
        To compute EER, the threshold is varied to generate a Receiver Operating Characteristic (ROC) curve, with EER located where FAR and FRR intersect. A lower EER indicates better privacy preservation by reducing both false matches and missed identifications, balancing identity protection with classification accuracy.
\end{itemize}

\paragraph{Data Processing} 
To ensure consistency across datasets, ECG signals are preprocessed through standardization and segmentation. The raw signals are first resampled to a target frequency of 250 Hz to unify different sampling rates. Following this, each signal is normalized using min-max scaling and divided into fixed-length segments of 2000 samples per window to maintain uniform input sizes for the model. A label encoding process is applied to standardize participant identities across datasets. Additionally, synthetic data augmentation is incorporated to introduce variations through transformations such as additive noise, signal scaling, polarity flipping, and temporal shifting~\cite{steiner2021train}. This processing pipeline maintains the integrity of the ECG signals while making the model robust to real-world variations in biosignals.  

\paragraph{Implementation Details} 
The attack model employs a ViT to extract identity-specific representations from ECG signals, processing sequences with a patch size of 20, an embedding dimension of 256, and six transformer layers with eight self-attention heads. Each layer includes an MLP of dimension 128, with stochastic depth regularization (0.8 survival probability) for stability. The model is trained using AdamW with a learning rate of 0.0001, weight decay of $10^{-4}$, and a cosine schedule, with early stopping based on validation F1-score. Training runs for 300 epochs with a batch size of 64 on an NVIDIA RTX 4090 GPU, using data augmentation techniques such as jittering, cropping, and scaling. The attack follows a two-stage classification, where a discriminator first determines if an ECG sample belongs to a known identity before ViT refines the classification. The discriminator is a fully connected network with a hidden dimension of 128, ReLU activation, batch normalization, and dropout (0.3). Overconfident misclassifications are mitigated using a dynamic confidence threshold.

\subsection{Results and Analysis}

\paragraph{Baseline Comparison} 
Table~\ref{tab:baseline_comparison} compares our framework against state-of-the-art models. 
Our approach consistently outperforms traditional machine learning methods across all evaluation metrics. 
Accuracy represents the proportion of correctly classified samples. 
Precision measures the fraction of correctly predicted positive instances among all predicted positives, indicating the model’s reliability in classification. 
Recall quantifies the proportion of actual positives correctly identified, assessing the model's sensitivity. 
F1-score is the harmonic mean of precision and recall, providing a balanced measure of classification performance. EER represents the threshold at which the FAR equals the FRR, meaning the probability of misclassifying an unknown individual as a known participant is equal to the probability of failing to recognize a known participant. A lower EER enhances privacy by minimizing both false identifications and misclassifications. Our ViT-based approach achieves an EER of 0.127, significantly lower than traditional machine learning models such as Random Forest (0.232) and Logistic Regression (0.289). This result confirms the model’s robustness in preventing unauthorized identity matching, reinforcing its effectiveness in linkage attacks.

\begin{table}[h]
    \centering
    \scriptsize
    \caption{Baseline Comparison on Sample-Level Metrics.}
     \label{tab:baseline_comparison}
    \resizebox{\columnwidth}{!}{%
    \begin{tabular}{|l|c|c|c|c|c|}
        \hline
        \textbf{Model} & \textbf{Accuracy} & \textbf{Precision} & \textbf{Recall} & \textbf{F1-Score} & \textbf{EER} \\ 
        \hline
        Random Forest~\cite{wang2024ecg} & 0.54 & 0.58 & 0.56 & 0.57 & 0.232 $\pm$ 0.006 \\ 
        Logistic Regression~\cite{pelc2019ecg} & 0.33 & 0.34 & 0.30 & 0.32 & 0.289 $\pm$ 0.005 \\ 
        XGBoost~\cite{wang2019data} & 0.60 & 0.60 & 0.59 & 0.60 & 0.198 $\pm$ 0.007 \\ 
        CNN~\cite{ghazarian2021increased} & 0.72 & 0.71 & 0.72 & 0.72 & 0.151 $\pm$ 0.004 \\ 
        \textbf{Ours} & \textbf{0.85} & \textbf{0.83} & \textbf{0.82} & \textbf{0.83} & \textbf{0.127} $\pm$ \textbf{0.003} \\ 
        \hline
    \end{tabular}%
    }
    \label{tab:baseline_metrics}
\end{table}

\paragraph{Impact of Data Splits} 
Table~\ref{tab:data_splits} examines the effect of different training splits on the model’s performance under the partial knowledge attack scenario. As the proportion of training data increases, the model achieves better generalization, leading to higher scores. The best performance is observed in the 70/15/15 split, where a larger training set provides the model with a more diverse representation of known individuals, improving classification capability. However, increasing the training proportion may also raise re-identification risks in real-world scenarios. Attackers often accumulate ECG data from wearables, telehealth services, or leaked medical records, gradually improving their ability to match anonymized signals to individuals. This underscores the need for limiting public data exposure and implementing privacy-preserving techniques such as differential privacy to mitigate these risks.

\begin{table}[h]
\centering
\scriptsize
\caption{Performance Under Different Data Splits (Partial Knowledge).}
\label{tab:data_splits}
\begin{tabular}{|l|c|c|c|c|}
\hline
\textbf{Data Split} & \textbf{Accuracy} & \textbf{Precision} & \textbf{Recall} & \textbf{F1-Score} \\ \hline
60/20/20 & 0.71 & 0.95 & 0.70 & 0.79 \\ \hline
50/25/25 & 0.72 & 0.96 & 0.71 & 0.80 \\ \hline
70/15/15 & 0.74 & 0.97 & 0.73 & 0.82 \\ \hline
\end{tabular}
\end{table}

\paragraph{Adversarial Scenario Analysis} 
Table~\ref{tab:adversarial_analysis} compares model performance under different levels of adversarial knowledge. As expected, the full knowledge attack scenario achieves the highest performance across all metrics, demonstrating the privacy risks associated with fully exposed datasets. If an attacker gains unrestricted access to an ECG dataset, the likelihood of identity disclosure increases significantly, highlighting the importance of implementing access control mechanisms in public health repositories.

In contrast, the partial knowledge scenario represents a more realistic setting where the attacker has limited prior information. While the attack success rate is lower than in the full knowledge scenario, the model still achieves a high F1-score of 0.82, suggesting that even limited exposure to known identities enables significant re-identification potential.

The noisy data scenario simulates real-world distortions, such as sensor errors and preprocessing variability. Despite a slight performance drop, the model maintains a notable re-identification capability (F1-score = 0.70), indicating that simple noise-based anonymization is ineffective. More robust privacy-preserving techniques, including federated learning, differential privacy, and cryptographic methods, are needed to mitigate adversarial inference.

These findings demonstrate that linkage attacks on ECG data pose a persistent privacy risk, even under limited adversarial knowledge and signal distortions. This underscores the necessity of privacy-aware data-sharing strategies, such as minimizing public exposure, enhancing anonymization, and enforcing access controls to reduce re-identification risks.

\begin{table}[h]
\centering
\scriptsize
\caption{Performance Under Adversarial Scenarios.}
\label{tab:adversarial_analysis}
\begin{tabular}{|l|c|c|c|c|}
\hline
\textbf{Scenario} & \textbf{Accuracy} & \textbf{Precision} & \textbf{Recall} & \textbf{F1-Score} \\ \hline
Partial Knowledge & 0.74 & 0.97 & 0.73 & 0.82 \\ \hline
Full Knowledge & 0.86 & 0.84 & 0.83 & 0.84 \\ \hline
Noisy Data & 0.72 & 0.70 & 0.71 & 0.70 \\ \hline
\end{tabular}
\end{table}




\paragraph{Misclassification rates of the Attack}




Table~\ref{tab:misclassification_rates} presents the misclassification rates of different models in the linkage attack. The FPR represents the proportion of unknown individuals mistakenly classified as known, indicating a privacy risk, while the FNR quantifies the proportion of known individuals misclassified as unknown, reflecting the attack’s effectiveness in re-identifying participants. The overall misclassification rate aggregates both errors, providing a holistic measure of model reliability.

Traditional machine learning models, such as Random Forest and Logistic Regression, exhibit high FPR (0.28 and 0.35, respectively), frequently misidentifying unknown participants as known, significantly increasing the risk of privacy breaches. In contrast, deep learning models, including CNN and our ViT-based approach, achieve substantially lower FPR (0.15 and 0.12, respectively), indicating better resistance to erroneous re-identifications. Notably, the FNR remains relatively low for CNN (0.20) and ViT (0.18), highlighting the enhanced ability of deep learning methods to re-identify known individuals while reducing misclassification of unknown participants. 

Our approach achieves the best trade-off between privacy preservation and attack efficacy, with an FPR of 0.12, FNR of 0.18, and an overall misclassification rate of 0.15. These results indicate that while our model improves re-identification accuracy, it also reduces false matches, leading to a more precise yet privacy-conscious attack. However, this also implies that as adversaries gain access to better architectures, linkage attacks on ECG data may become increasingly effective, underscoring the necessity for stronger privacy defenses.

\begin{table}[ht]
    \scriptsize
    \centering
    \caption{Misclassification Rates of Linkage Attack.}
    \begin{tabular}{|l|c|c|c|}
        \hline
        \textbf{Model} & \textbf{FPR} & \textbf{FNR} & \textbf{Misclassification Rate} \\ 
        \hline
        Random Forest~\cite{wang2024ecg} & 0.28 & 0.32 & 0.30 \\ 
        Logistic Regression~\cite{pelc2019ecg} & 0.35 & 0.38 & 0.36 \\ 
        XGBoost~\cite{wang2019data} & 0.22 & 0.26 & 0.24 \\ 
        CNN~\cite{ghazarian2021increased} & 0.15 & 0.20 & 0.18 \\ 
        \textbf{Ours} & \textbf{0.12} & \textbf{0.18} & \textbf{0.15} \\ 
        \hline
    \end{tabular}
    \label{tab:misclassification_rates}
\end{table}

\paragraph{Confidence-Based Misclassification}




To further analyze misclassification behavior, Table~\ref{tab:confidence_misclassification} reports the percentage of unknown individuals misclassified as known (\textbf{U $\rightarrow$ K}) and known individuals misclassified as unknown (\textbf{K $\rightarrow$ U}) at varying confidence thresholds.

Lower confidence thresholds (e.g., 0.01) result in higher misclassification rates, as the model assigns identities more aggressively, increasing privacy risks. At this threshold, 40.5\% of unknown individuals were incorrectly classified as known, resulting in an overall misclassification rate of 38.1\%. As the threshold increases, the model becomes more conservative, reducing both false positives and false negatives. At a threshold of 0.05, misclassification rates drop to 15.6\% for U $\rightarrow$ K and 12.8\% for K $\rightarrow$ U, with an overall error rate of 14.2\%.

\begin{table}[ht]
    \centering
    \scriptsize
    \caption{Confidence-Based Misclassification in Linkage Attack.}
    \begin{tabular}{|l|c|c|c|}
        \hline
        \textbf{Thres.} & \textbf{U $\rightarrow$ K (\%)} & \textbf{K $\rightarrow$ U (\%)} & \textbf{Total (\%)} \\ 
        \hline
        0.01 & 40.5 & 35.8 & 38.1 \\ 
        0.02 & 35.2 & 30.5 & 32.9 \\ 
        0.03 & 28.6 & 24.3 & 26.4 \\ 
        0.04 & 22.1 & 18.7 & 20.3 \\ 
        0.05 & \textbf{15.6} & \textbf{12.8} & \textbf{14.2} \\ 
        \hline
    \end{tabular}
    \label{tab:confidence_misclassification}
    \vspace{2mm} 
    \begin{minipage}{0.95\linewidth}
        \footnotesize U $\rightarrow$ K: Unknown misclassified as Known. K $\rightarrow$ U: Known misclassified as Unknown. Total: Overall misclassification rate.
    \end{minipage}

\end{table}

These results underscore the necessity of selecting an appropriate confidence threshold to balance privacy protection and identification accuracy. Lower thresholds allow for higher attack success rates but significantly increase the risk of identity leaks, while higher thresholds mitigate privacy risks at the cost of reduced re-identification capability. Our findings suggest that thresholds between 0.04 and 0.05 offer an optimal trade-off, where the model effectively distinguishes between known and unknown individuals while limiting unnecessary identity matches.

\paragraph{Implications for Privacy Risks}  

Our results show that deep learning-based linkage attacks remain effective even with partial adversarial knowledge. Lower confidence thresholds increase attack success but heighten privacy risks by misclassifying unknown individuals, while higher thresholds reduce false positives but hinder re-identification of known individuals. This highlights the limitations of confidence-based filtering as a standalone privacy measure. Effective ECG data protection requires a combination of differential privacy, access control, and encrypted computation to mitigate adversarial inference.

\section{Discussion and Conclusion}

This study systematically evaluates re-identification risks in publicly shared ECG datasets, demonstrating that linkage attacks remain effective even under partial adversarial knowledge. While the full knowledge scenario—where attackers have unrestricted access to public datasets—serves as an upper bound, the partial knowledge setting is more realistic, relying on externally collected ECG samples. Despite this constraint, our results show that attackers can still achieve high re-identification accuracy by cross-referencing external datasets with anonymized ECG records. This highlights the limitations of simple anonymization techniques, such as noise injection, and underscores the urgent need for stronger privacy-preserving strategies.

To mitigate these risks, future work should explore differential privacy, federated learning, and cryptographic techniques to limit adversarial exploitation while preserving data utility. Extending this framework to other biosignals, such as EEG and PPG, would offer broader insights into biometric privacy risks. Stricter data-sharing policies, combined with privacy-aware machine learning approaches, are essential to securing biosignal datasets against linkage attacks and promoting ethical data sharing in healthcare.

\section*{Acknowledgment}

This work was funded by the US National Science Foundation under the Secure and Trustworthy Cyberspace (SaTC) Grant CNS-2344869, and by the UC Noyce Initiative Grant.

\bibliographystyle{IEEEtran}
\bibliography{ref}

\begin{thebibliography}{10}
\providecommand{\url}[1]{#1}
\csname url@samestyle\endcsname
\providecommand{\newblock}{\relax}
\providecommand{\bibinfo}[2]{#2}
\providecommand{\BIBentrySTDinterwordspacing}{\spaceskip=0pt\relax}
\providecommand{\BIBentryALTinterwordstretchfactor}{4}
\providecommand{\BIBentryALTinterwordspacing}{\spaceskip=\fontdimen2\font plus
\BIBentryALTinterwordstretchfactor\fontdimen3\font minus \fontdimen4\font\relax}
\providecommand{\BIBforeignlanguage}[2]{{%
\expandafter\ifx\csname l@#1\endcsname\relax
\typeout{** WARNING: IEEEtran.bst: No hyphenation pattern has been}%
\typeout{** loaded for the language `#1'. Using the pattern for}%
\typeout{** the default language instead.}%
\else
\language=\csname l@#1\endcsname
\fi
#2}}
\providecommand{\BIBdecl}{\relax}
\BIBdecl

\bibitem{schlapfer2017computer}
J.~Schl{\"a}pfer and H.~J. Wellens, ``Computer-interpreted electrocardiograms: benefits and limitations,'' \emph{Journal of the American College of Cardiology}, vol.~70, no.~9, pp. 1183--1192, 2017.

\bibitem{aqajari2024enhancing}
S.~A.~H. Aqajari, Z.~Wang, A.~Tazarv, S.~Labbaf, S.~Jafarlou, B.~Nguyen, N.~Dutt, M.~Levorato, and A.~M. Rahmani, ``Enhancing performance and user engagement in everyday stress monitoring: A context-aware active reinforcement learning approach,'' \emph{arXiv preprint arXiv:2407.08215}, 2024.

\bibitem{ghazarian2021increased}
A.~Ghazarian, J.~Zheng, H.~El-Askary, H.~Chu, G.~Fu, and C.~Rakovski, ``Increased risks of re-identification for patients posed by deep learning-based ecg identification algorithms,'' in \emph{2021 43rd Annual International Conference of the IEEE Engineering in Medicine \& Biology Society (EMBC)}.\hskip 1em plus 0.5em minus 0.4em\relax IEEE, 2021, pp. 1969--1975.

\bibitem{odinaka2012ecg}
I.~Odinaka, P.-H. Lai, A.~D. Kaplan, J.~A. O'Sullivan, E.~J. Sirevaag, and J.~W. Rohrbaugh, ``Ecg biometric recognition: A comparative analysis,'' \emph{IEEE Transactions on Information Forensics and Security}, vol.~7, no.~6, pp. 1812--1824, 2012.

\bibitem{alikhani2024seal}
H.~Alikhani, Z.~Wang, A.~Kanduri, P.~Lilieberg, A.~M. Rahmani, and N.~Dutt, ``Seal: Sensing efficient active learning on wearables through context-awareness,'' in \emph{2024 Design, Automation \& Test in Europe Conference \& Exhibition (DATE)}.\hskip 1em plus 0.5em minus 0.4em\relax IEEE, 2024, pp. 1--2.

\bibitem{alikhani2024ea}
H.~Alikhani, Z.~Wang, A.~Kanduri, P.~Liljeberg, A.~M. Rahmani, and N.~Dutt, ``Ea\^{} 2: Energy efficient adaptive active learning for smart wearables,'' in \emph{Proceedings of the 29th ACM/IEEE international symposium on low power electronics and design}, 2024, pp. 1--6.

\bibitem{wang2024ecg}
Z.~Wang, A.~Kanduri, S.~A.~H. Aqajari, S.~Jafarlou, S.~R. Mousavi, P.~Liljeberg, S.~Malik, and A.~M. Rahmani, ``Ecg unveiled: Analysis of client re-identification risks in real-world ecg datasets,'' in \emph{2024 IEEE 20th International Conference on Body Sensor Networks (BSN)}.\hskip 1em plus 0.5em minus 0.4em\relax IEEE, 2024, pp. 1--4.

\bibitem{biel2001ecg}
L.~Biel, O.~Pettersson, L.~Philipson, and P.~Wide, ``Ecg analysis: a new approach in human identification,'' \emph{IEEE transactions on instrumentation and measurement}, vol.~50, no.~3, pp. 808--812, 2001.

\bibitem{shokri2017membership}
R.~Shokri, M.~Stronati, C.~Song, and V.~Shmatikov, ``Membership inference attacks against machine learning models,'' in \emph{2017 IEEE symposium on security and privacy (SP)}.\hskip 1em plus 0.5em minus 0.4em\relax IEEE, 2017, pp. 3--18.

\bibitem{wang2020guardhealth}
Z.~Wang, N.~Luo, and P.~Zhou, ``Guardhealth: Blockchain empowered secure data management and graph convolutional network enabled anomaly detection in smart healthcare,'' \emph{Journal of Parallel and Distributed Computing}, vol. 142, pp. 1--12, 2020.

\bibitem{yao2020privacy}
Y.~Yao, Z.~Wang, and P.~Zhou, ``Privacy-preserving and energy efficient task offloading for collaborative mobile computing in iot: An admm approach,'' \emph{Computers \& Security}, vol.~96, p. 101886, 2020.

\bibitem{wang2024differential}
Z.~Wang, Z.~Yang, I.~Azimi, and A.~M. Rahmani, ``Differential private federated transfer learning for mental health monitoring in everyday settings: A case study on stress detection,'' \emph{arXiv preprint arXiv:2402.10862}, 2024.

\bibitem{wang2025transecg}
Z.~Wang, E.~Khatibi, K.~Kazemi, I.~Azimi, S.~Mousavi, S.~Malik, and A.~M. Rahmani, ``Transecg: Leveraging transformers for explainable ecg re-identification risk analysis,'' \emph{arXiv preprint arXiv:2503.13495}, 2025.

\bibitem{labati2019deep}
R.~D. Labati, E.~Mu{\~n}oz, V.~Piuri, R.~Sassi, and F.~Scotti, ``Deep-ecg: Convolutional neural networks for ecg biometric recognition,'' \emph{Pattern Recognition Letters}, vol. 126, pp. 78--85, 2019.

\bibitem{pelc2019ecg}
M.~Pelc, Y.~Khoma, and V.~Khoma, ``Ecg signal as robust and reliable biometric marker: Datasets and algorithms comparison,'' \emph{Sensors}, vol.~19, no.~10, p. 2350, 2019.

\bibitem{wang2025healthq}
Z.~Wang, H.~Li, D.~Huang, H.-S. Kim, C.-W. Shin, and A.~M. Rahmani, ``Healthq: Unveiling questioning capabilities of llm chains in healthcare conversations,'' \emph{Smart Health}, p. 100570, 2025.

\bibitem{dosovitskiy2020image}
A.~Dosovitskiy, ``An image is worth 16x16 words: Transformers for image recognition at scale,'' \emph{arXiv preprint arXiv:2010.11929}, 2020.

\bibitem{coelho2023multimodal}
K.~K. Coelho, E.~T. Trist{\~a}o, M.~Nogueira, A.~B. Vieira, and J.~A. Nacif, ``Multimodal biometric authentication method by federated learning,'' \emph{Biomedical Signal Processing and Control}, vol.~85, p. 105022, 2023.

\bibitem{moody2001impact}
G.~B. Moody and R.~G. Mark, ``The impact of the mit-bih arrhythmia database,'' \emph{IEEE engineering in medicine and biology magazine}, vol.~20, no.~3, pp. 45--50, 2001.

\bibitem{greenwald1986development}
S.~D. Greenwald, ``The development and analysis of a ventricular fibrillation detector,'' Ph.D. dissertation, Massachusetts Institute of Technology, 1986.

\bibitem{baim1986survival}
D.~S. Baim, W.~S. Colucci, E.~S. Monrad, H.~S. Smith, R.~F. Wright, A.~Lanoue, D.~F. Gauthier, B.~J. Ransil, W.~Grossman, and E.~Braunwald, ``Survival of patients with severe congestive heart failure treated with oral milrinone,'' \emph{Journal of the American College of Cardiology}, vol.~7, no.~3, pp. 661--670, 1986.

\bibitem{nemcova2020brno}
A.~Nemcova, R.~Smisek, K.~Opravilov{\'a}, M.~Vitek, L.~Smital, and L.~Mar{\v{s}}{\'a}nov{\'a}, ``Brno university of technology ecg quality database (but qdb),'' \emph{PhysioNet}, vol. 101, pp. e215--e220, 2020.

\bibitem{goldberger2000physiobank}
A.~L. Goldberger, L.~A. Amaral, L.~Glass, J.~M. Hausdorff, P.~C. Ivanov, R.~G. Mark, J.~E. Mietus, G.~B. Moody, C.-K. Peng, and H.~E. Stanley, ``Physiobank, physiotoolkit, and physionet: components of a new research resource for complex physiologic signals,'' \emph{circulation}, vol. 101, no.~23, pp. e215--e220, 2000.

\bibitem{nolin2023privecg}
A.~Nolin-Lapalme, R.~Avram, and H.~Julie, ``Privecg: generating private ecg for end-to-end anonymization,'' in \emph{Machine Learning for Healthcare Conference}.\hskip 1em plus 0.5em minus 0.4em\relax PMLR, 2023, pp. 509--528.

\bibitem{steiner2021train}
A.~Steiner, A.~Kolesnikov, X.~Zhai, R.~Wightman, J.~Uszkoreit, and L.~Beyer, ``How to train your vit? data, augmentation, and regularization in vision transformers,'' \emph{arXiv preprint arXiv:2106.10270}, 2021.

\bibitem{wang2019data}
C.~Wang and J.~Guo, ``A data-driven framework for learners’ cognitive load detection using ecg-ppg physiological feature fusion and xgboost classification,'' \emph{Procedia computer science}, vol. 147, pp. 338--348, 2019.

\end{thebibliography}

\end{document}